\documentclass[preprint,amsmath,amssymb,aps,pra,longbibliography]{revtex4-1}

\usepackage[margin=0.75in, top=1in, bottom=0.85in]{geometry}

\usepackage[final,letterspace=-0]{microtype}
\usepackage[colorlinks=true, linkcolor=blue, urlcolor=blue, citecolor=blue, anchorcolor=blue]{hyperref}

\usepackage{mathpazo} % add possibly `sc` and `osf` options
\usepackage{upgreek}

\usepackage{graphicx}% Include figure files
\usepackage{dcolumn}% Align table columns on decimal point
\usepackage{bm}% bold math
\usepackage{helvet}

\begin{document}

\title{Observation of optical de~Broglie-Mackinnon wave packets}

\author{Layton A. Hall$^{1}$ and Ayman F. Abouraddy$^{1,}$}
\email{Corresponding authors: raddy@creol.ucf.edu}
\affiliation{$^{1}$CREOL, The College of Optics \& Photonics, University of Central Florida, Orlando, Florida 32816, USA}

\begin{abstract}
de~Broglie wave packets accompanying moving particles are dispersive and lack an intrinsic length scale dictated solely by the particle mass and velocity. Mackinnon proposed almost 45~years ago a localized non-dispersive wave packet constructed out of dispersive de~Broglie phase waves via a Copernican inversion of the roles of particle and observer, whereupon an intrinsic length scale emerges by accounting for every possible observer -- rather than by introducing an \textit{ad hoc} uncertainty in the particle velocity. The de~Broglie-Mackinnon (dBM) wave packet has nevertheless remained to date a theoretical entity. Here, we report the observation of optical dBM wave packets using paraxial space-time-coupled pulsed laser fields in presence of anomalous group-velocity dispersion. Crucially, the bandwidth of dBM wave packets has an upper limit that is compatible with the wave-packet group velocity and equivalent mass. In contrast to previously observed linear propagation-invariant wave packets whose spatio-temporal profiles at any axial plane are X-shaped, those for dBM wave packets are uniquely O-shaped (circularly symmetric with respect to space and time). By sculpting their spatio-temporal spectral structure, we produce dispersion-free dBM wave packets in the dispersive medium, observe their circularly symmetric spatio-temporal profiles, and tune the field parameters corresponding to particle mass and velocity that uniquely determine the wave-packet length scale.
\end{abstract}

\maketitle

%\noindent

It is well-known that there are no dispersion-free wave-packet solutions to the $(1+1)$D potential-free Schr{\"o}dinger equation -- with the sole exception of the Airy wave packet identified by Berry and Balasz in 1979 \cite{Berry79AMP}, which does \textit{not} travel at a fixed group velocity, but rather accelerates despite the absence of an external force \cite{Greenberger80AJP}. The Airy wave packet has impacted all areas of wave physics (e.g., optics \cite{Siviloglou07OL}, acoustics \cite{Zhang14NC}, water waves \cite{Fu2015PRL}, electron beams \cite{Bloch13Nature}, and as a model for Dirac particles \cite{Kaminer15NPhys}). Less-known is that in the year preceding the discovery of the Airy wave packet, Mackinnon identified a non-dispersive $(1+1)$D wave packet that travels at a \textit{constant} group velocity \cite{Mackinnon78FP}, but is constructed out of dispersive de~Broglie `phase waves' that accompany the motion of a massive particle and are solutions to the Klein-Gordon equation. de~Broglie had originally demonstrated that the group velocity $\widetilde{v}$ of a wave packet constructed of phase-waves is equal to the particle velocity $v$ \cite{deBroglie25}. However, localized de~Broglie wave packets are dispersive, as are Schr{\"o}dinger wave packets \cite{TannoudjiBook}. Moreover, because de~Broglie wave packets necessitate introducing an \textit{ad hoc} uncertainty in the particle velocity \cite{deBroglie25}, and there is no upper limit on the exploitable bandwidth, such wave packets lack an \textit{intrinsic} length scale (i.e., a scale uniquely determined by the particle mass and velocity).

Through a Copernican inversion of the roles of particle and observer, Mackinnon constructed out of dispersive de~Broglie phase waves a \textit{non-dispersive} wave packet \cite{Mackinnon78FP} -- which we refer to henceforth as the de~Broglie-Mackinnon (dBM) wave packet. Instead of introducing an \textit{ad hoc} uncertainty into the particle velocity from the perspective of a privileged reference frame, Mackinnon suggested accounting for all possible observers, who cooperatively report observations made in their reference frames to a single agreed-upon frame in which Lorentz contraction and time dilation are corrected for \cite{Mackinnon78FP}. Besides retaining the salutary features of conventional de~Broglie wave packets, Mackinnon's construction unveiled an intrinsic length scale for the dBM wave packet determined solely by the particle mass and velocity. However, despite the clear algorithmic process for constructing the dBM wave packet, it is \textit{not} a solution to the Klein-Gordon equation \cite{Mackinnon78FP}, and is instead constructed only epistemologically in the selected reference frame. As such, dBM wave packets have yet to be realized in any physical wave.

Nevertheless, it has been recognized that the $(1+1)$D dBM wave packet can be mapped to physical solutions of the optical wave equation by first enlarging the field dimensionality to $(2+1)$D, which  allows introducing angular dispersion \cite{Torres10AOP,Fulop10Applications}. This procedure enables realizing the dBM dispersion relationship for propagation along the optical axis in the initial reduced-dimensionality $(1+1)$D space \cite{Saari04PRE}. However, observing optical dBM wave packets in free space faces insurmountable practical difficulties \cite{ZamboniRached2008PRA,Yessenov22AOP}. Specifically, such wave packets are produced by relativistic optical dipoles and are observed by stationary, coherent field detectors that nevertheless fully encircle the moving dipole. We investigate here a different strategy that makes use of the unique characteristics of optical-wave propagation in the anomalous group-velocity dispersion (GVD) regime \cite{SalehBook07} to produce paraxial dBM wave packets. In this conception, an optical dBM wave packet is a particular realization of so-called `space-time' (ST) wave packets \cite{Kondakci16OE,Parker16OE,Wong17ACSP2,Yessenov19OPN,Yessenov22AOP} in dispersive media \cite{Porras03OL,Porras03PRE2,Longhi04OL,Porras04PRE,Malaguti08OL,Malaguti09PRA,Hall22LPR}. In general, ST wave packets are pulsed optical beams whose unique characteristics (e.g., tunable group velocity \cite{Kondakci19NC} and anomalous refraction \cite{Bhaduri20NatPhot}) stem from their spatio-temporal spectral structure rather than their particular spatial or temporal profiles. Recent advances in the synthesis of ST wave packets make them a convenient platform for producing a wide variety of structured pulsed fields \cite{Yessenov22AOP}, including dBM wave packets.

Here, we provide unambiguous observations of optical dBM wave packets in presence of anomalous GVD. Starting with generic femtosecond pulses, we make use of a universal angular-dispersion synthesizer \cite{Hall21OE2} to construct spatio-temporally structured optical fields in which the spatial and temporal degrees-of-freedom are no longer separable. Critically, the association between the propagation angle and wavelength is two-to-one rather than one-to-one as in conventional tilted pulse fronts \cite{Torres10AOP}. This feature allows folding the spatio-temporal spectrum back on itself, thereby guaranteeing the paraxiality of the synthesized dBM wave packets. Consequently, these wave packets retain in the medium all the characteristic features of their free-space counterparts while circumventing the above-mentioned difficulties. Such space-time-coupled wave packets are dispersive in free space, but become propagation-invariant once coupled to a medium in the anomalous-GVD regime, where they travel at a tunable group velocity $\widetilde{v}$. Although all previously observed linear, propagation-invariant wave packets have at a fixed axial plane been either X-shaped \cite{Saari97PRL,Turunen10PO,FigueroaBook14,Yessenov22AOP} or separable \cite{Chong10NP} with respect to the transverse coordinate and time, the spatio-temporal profiles of dBM wave packets are -- in contrast -- circularly symmetric (O-shaped). In addition to verifying this long-theorized O-shaped spatio-temporal structure \cite{Longhi04OL,Porras04PRE,Malaguti08OL}, we confirm the impact of the two identifying parameters (equivalent to particle mass and velocity) on the bandwidth and length scale of the non-dispersive dBM wave packets. Propagation invariance in the dispersive medium constrains the maximum bandwidth (minimum wave-packet length) according to these selected parameters. Finally, in contrast to Airy wave packets that are the unique non-dispersive solution to the Schr{\"o}dinger equation, the axial profile of dBM wave packets can be varied almost arbitrarily, which we confirm by modulating their spatio-temporal spectral phase distribution. These results may pave the way to optical tests of the solutions of the Klein-Gordon equation for massive particles.

%\clearpage

\section*{Results}

\subsection*{Theory of de~Broglie wave packets}

\noindent
de~Broglie posited two distinct entities accompanying massive particles: an \textit{internal} `clock' and an \textit{external} `phase wave' \cite{MacKinnon76AJP,Espinosa82AJP}. For a particle of rest mass $m_{\mathrm{o}}$ whose energy is expressed as $E_{\mathrm{o}}\!=\!m_{\mathrm{o}}c^{2}\!=\!\hbar\omega_{\mathrm{o}}$, the internal clock and the infinite-wavelength phase wave coincide at the same de~Broglie frequency $\omega_{\mathrm{o}}$ in the particle's rest frame [Fig.~\ref{Fig:ConceptofDBM}(a)]; here $c$ is the speed of light in vacuum, and $\hbar$ is the modified Planck constant. When the particle moves at a velocity $v$, the frequencies observed in the rest frame diverge: the internal frequency drops to $\omega\!=\!\omega_{\mathrm{o}}\sqrt{1-\beta_{v}^{2}}$ whereas the phase-wave frequency increases to $\omega\!=\!\omega_{\mathrm{o}}\big/\sqrt{1-\beta_{v}^{2}}$ and takes on a finite wavelength $\lambda$, where $\beta_{v}\!=\!\tfrac{v}{c}$ [Fig.~\ref{Fig:ConceptofDBM}(b)]. The wave number $k\!=\!\tfrac{2\pi}{\lambda}$ for the phase wave is determined by the de~Broglie dispersion relationship $\omega^{2}\!=\!\omega_{\mathrm{o}}^{2}+c^{2}k^{2}$ [Fig.~\ref{Fig:ConceptofDBM}(c)], so that it is a solution to the Klein-Gordon equation. Because de~Broglie phase waves are extended, a particle with a well-defined velocity cannot be localized. Instead, spatially localizing the particle requires introducing an \textit{ad hoc} uncertainty in the particle velocity (a spread from $v$ to $v+\Delta v$) to induce a bandwidth $\Delta\omega$ (from $\omega_{\mathrm{c}}$ to $\omega_{\mathrm{c}}+\Delta\omega$), or $\Delta k$ (from $k_{\mathrm{c}}$ to $k_{\mathrm{c}}+\Delta k$) \cite{deBroglie25,Mackinnon78FP}, thus resulting in a finite-width wave packet that is also a solution to the Klein-Gordon equation [Fig.~\ref{Fig:ConceptofDBM}(c)]. The wave-packet \textit{group velocity} $\widetilde{v}\!=\!1\big/\tfrac{dk}{d\omega}\big|_{\omega_{\mathrm{c}}}\!=\!v$ is equal to the particle velocity, whereas its phase velocity is $v_{\mathrm{ph}}\!=\!\tfrac{\omega}{k}\!=\!\tfrac{c^{2}}{v}$  ($v_{\mathrm{ph}}\widetilde{v}\!=\!c^{2}$; see Methods). However, de~Broglie wave packets are dispersive $\tfrac{d\widetilde{v}}{d\omega}\!\neq\!0$. Moreover, because there is no upper limit on the exploitable bandwidth [Fig.~\ref{Fig:ConceptofDBM}(c)], de~Broglie wave packets lack an intrinsic length scale; that is, there is no \textit{minimum} wave-packet length that is uniquely identified by the particle parameters (mass $m_{\mathrm{o}}$ and velocity $v$).

\subsection*{Non-dispersive de~Broglie-Mackinnon (dBM) wave packets}

\noindent
Mackinnon proposed an altogether different conception for constructing localized \textit{non-dispersive} wave packets out of de~Broglie phase waves that jettisons the need for introducing an \textit{ad hoc} uncertainty in particle velocity to localize it. Key to this proposal is a Copernican inversion of the roles of particle and observer. Rather than a single privileged observer associated with the rest frame in Fig.~\ref{Fig:ConceptofDBM}(c), Mackinnon considered a continuum of potential observers traveling at physically accessible velocities (from $-c$ to $c$). The wave-packet bandwidth $\Delta k$ that is established in a particular reference frame is a result of the spread in the particle velocity as observed in all these possible frames. Consequently, the particle can be localized, and a unique wave-packet length scale identified, even when its velocity is well-defined.

The physical setting envisioned by Mackinnon is depicted in Fig.~\ref{Fig:ConceptofDBM}(d), where the particle moves at a velocity $v$ and an observer moves at $u$, both with respect to a common rest frame in which the dBM wave packet is constructed. Each potential observer records a different phase-wave frequency and wavelength. The crucial step is that \textit{all} potential observers travelling at velocities $u$ ranging from $-c$ to $c$ report their observations to the selected rest frame. These phase waves are superposed in this frame -- after accounting for Lorentz contraction and time dilation (Methods) -- to yield a wave packet uniquely identified by the particle rest mass $m_{\mathrm{o}}$ and velocity $v$. 

Consider first the simple scenario where the particle is at rest with respect to the selected frame ($v\!=\!0$). Each observer reports to the common rest frame a frequency $\omega'\!=\!\omega_{\mathrm{o}}/\sqrt{1-\beta_{u}^{2}}$ and a wave number $k'\!=\!-k_{\mathrm{o}}\beta_{u}/\sqrt{1-\beta_{u}^{2}}$, where $\beta_{u}\!=\!\tfrac{u}{c}$. Accounting for time dilation results in $\omega'\rightarrow\omega\!=\!\omega_{\mathrm{o}}$, and accounting for Lorentz contraction produces $k'\rightarrow k\!=\!-k_{\mathrm{o}}\beta_{u}$. Therefore, the frequency in the rest frame based on the recordings of \textit{all} the observers is $\omega\!=\!\omega_{\mathrm{o}}$, just as in the case of a conventional de~Broglie phase wave, but the wave number now extends over the range from $-k_{\mathrm{o}}$ to $k_{\mathrm{o}}$ as the observer velocity $u$ ranges from $c$ to $-c$ [Fig.~\ref{Fig:ConceptofDBM}(e)]. In other words, the observer velocity $u$ serves as an internal parameter that is swept to establish a new dispersion relationship whose slope is zero, thus indicating a particle at rest $\widetilde{v}\!=\!v\!=\!0$ \cite{Mackinnon78FP,Saari04PRE}. The spectral representation of the support domain for this wave packet is a horizontal line $\omega\!=\!\omega_{\mathrm{o}}$ in $(k,\tfrac{\omega}{c})$-space delimited by the two light-lines $k\!=\!\pm\tfrac{\omega}{c}$ [Fig.~\ref{Fig:ConceptofDBM}(e)]. In contradistinction to conventional de~Broglie wave packets, a physically motivated length scale emerges for the dBM wave packet. The maximum spatial bandwidth is $\Delta k\!=\!2k_{\mathrm{o}}$, which corresponds to a minimum wave-packet length scale of $L_{\mathrm{min}}\!\sim\!\tfrac{\lambda_{\mathrm{o}}}{2}$, where $\lambda_{\mathrm{o}}\!=\!\tfrac{2\pi}{k_{\mathrm{o}}}$. This can be viewed as an `optical theorem', whereby the dBM wave packet for a stationary particle cannot be spatially localized below the associated de~Broglie wavelength $\lambda_{\mathrm{o}}$. Taking an equal-weight superposition across all the wave numbers, the dBM wave packet is $\psi(z;t)\propto e^{-i\omega_{\mathrm{o}}t}\mathrm{sinc}(\tfrac{\Delta k}{\pi}z)$, where $\mathrm{sinc}(x)\!=\!\tfrac{\sin{\pi x}}{\pi x}$ \cite{Mackinnon78FP}.
 
A similar procedure can be followed when $v\!\neq\!0$, whereupon the frequency and wave number in the selected reference frame are $\omega\!=\!\omega_{\mathrm{o}}(1-\beta_{v}\beta_{u})\big/\sqrt{1-\beta_{v}^{2}}$ and $k\!=\!k_{\mathrm{o}}(\beta_{v}-\beta_{u})\big/\sqrt{1-\beta_{v}^{2}}$, respectively (Methods). Because $v$ is fixed whereas $u$ extends from $-c$ to $c$, a linear dispersion relationship between $\omega$ and $k$ is established, $k\!=\!\tfrac{1}{\beta_{v}}(\tfrac{\omega}{c}-\tfrac{k_{\mathrm{o}}^{2}}{k_{1}})$, where $k_{1}\!=\!k_{\mathrm{o}}/\sqrt{1-\beta_{v}^{2}}$. The slope of the dBM dispersion relationship indicates that $\widetilde{v}\!=\!v$ as in conventional de~Broglie wave packets, but the dBM wave packet is now \textit{non-dispersive}, $\tfrac{d\widetilde{v}}{d\omega}\!=\!0$ [Fig.~\ref{Fig:ConceptofDBM}(f)]. The limits on the spatial and temporal bandwidths for the dBM wave packet are $\Delta k\!=\!2k_{1}$ and $\tfrac{\Delta\omega}{c}\!=\!\beta_{v}\Delta k$, respectively, leading to a reduced characteristic length scale $L_{\mathrm{min}}\!\sim\!\tfrac{\lambda_{\mathrm{o}}}{2}\sqrt{1-\beta_{v}^{2}}$ as a manifestation of Lorentz contraction; a faster particle is more tightly localized. By assigning equal complex amplitudes to all the phase waves associated with this moving particle, the propagation-invariant dBM wave packet is $\psi(z;t)\propto e^{i\beta_{v}\Delta k(z-\widetilde{v}t)}\mathrm{sinc}(\tfrac{\Delta k}{\pi}(z-\widetilde{v}t))$. Crucially, unlike conventional de~Broglie wave packets, the dBM wave packet is \textit{not} a solution to the Klein-Gordon equation, although a modified wave equation can perhaps be constructed for it \cite{Mackinnon78FP}.

%\clearpage
\subsection*{Optical de~Broglie-Mackinnon wave packets in free space}

\noindent
Despite their intrinsic interest from a fundamental point of view, dBM wave packets have remained to date theoretical entities. It has nevertheless been recognized that \textit{optical} waves in free space may provide a platform for their construction \cite{Saari04PRE,ZamboniRached2008PRA}. Because $(1+1)$D optical waves in free space are dispersion-free ($k\!=\!\tfrac{\omega}{c}$ and $v_{\mathrm{ph}}\!=\!\widetilde{v}\!=\!c$), producing optical dBM wave packets requires first adding a transverse coordinate $x$ to enlarge the field dimensionality to $(2+1)$D. The dispersion relationship thus becomes $k_{x}^{2}+k_{z}^{2}\!=\!(\tfrac{\omega}{c})^{2}$, which represents the surface of a `light-cone' \cite{Donnelly93ProcRSLA,Yessenov22AOP}; here $k_{x}$ and $k_{z}$ are the transverse and longitudinal components of the wave vector along $x$ and $z$, respectively. The spectral support of any optical field corresponds to some region on the light-cone surface [Fig.~\ref{Fig:OpticaldBMConcept}(a)]. For a fixed value of $k_{x}\!=\!\pm\tfrac{\omega_{\mathrm{o}}}{c}$, we retrieve the axial dispersion relationship for de~Broglie phase waves $\omega^{2}\!=\!\omega_{\mathrm{o}}^{2}+c^{2}k_{z}^{2}$. A convenient parametrization of the field makes use of the propagation angle $\varphi(\omega)$ with respect to the $z$-axis for the plane wave at a frequency $\omega$, whereupon $k_{x}(\omega)\!=\!\tfrac{\omega}{c}\sin{\varphi(\omega)}$ and $k_{z}(\omega)\!=\!\tfrac{\omega}{c}\cos{\varphi(\omega)}$. \textit{Angular dispersion} is thus introduced into the $(2+1)$D field \cite{Torres10AOP,Fulop10Applications}, and its spectral support on the light-cone surface is a one-dimensional (1D) trajectory. We take \textit{optical} dBM wave packets to be those whose axial dispersion relationship $\omega(k_{z})$ conforms to that of a dBM wave packet. This requires that the projection of the spectral support onto the $(k_{z},\tfrac{\omega}{c})$-plane be linear and delimited by the light-lines $k_{z}\!=\!\pm\tfrac{\omega}{c}$. Indeed, the spectral projections onto the $(k_{z},\tfrac{\omega}{c})$-plane in Fig.~\ref{Fig:OpticaldBMConcept}(a,b) coincide with those in Fig.~\ref{Fig:ConceptofDBM}(e,f).

Consider first a monochromatic field $\omega\!=\!\omega_{\mathrm{o}}$ whose spectral support is the circle at the intersection of the light-cone with a horizontal iso-frequency plane [Fig.~\ref{Fig:OpticaldBMConcept}(a)].  This monochromatic field comprises plane waves of the same frequency $\omega_{\mathrm{o}}$ that travel at angles $\varphi$ extending from 0 to $2\pi$, whose axial wave numbers are $k_{z}(\varphi)\!=\!\pm\sqrt{k_{\mathrm{o}}^{2}-k_{x}^{2}}\!=\!k_{\mathrm{o}}\cos\varphi$ and extend from $-k_{\mathrm{o}}$ to $k_{\mathrm{o}}$. This optical wave packet [Fig.~\ref{Fig:OpticaldBMConcept}(a)] corresponds to the dBM wave packet for a particle in its rest frame [Fig.~\ref{Fig:ConceptofDBM}(e)], and $\varphi$ serves as the new internal parameter to be swept in order to produce the targeted dBM dispersion relationship, corresponding to the observer velocity $u$ in Fig.~\ref{Fig:ConceptofDBM}(e). By setting the spectral amplitudes equal for all the plane-wave components, we obtain $\psi(x,z;t)\propto e^{-i\omega_{\mathrm{o}}t}\,\mathrm{sinc}(\tfrac{\Delta k_{z}}{\pi}\sqrt{x^{2}+z^{2}})$, where $\Delta k_{z}\!=\!2k_{\mathrm{o}}$ [Fig.~\ref{Fig:OpticaldBMConcept}(a)]. Such a wave packet can be produced by a stationary, monochromatic planar dipole placed at the origin of the $(x,z)$-plane. Observing this optical field requires coherent field detectors arranged around the $2\pi$ angle subtended by the dipole, and then communicating the recorded measurements to a central station. This procedure is therefore \textit{not} dissimilar in principle from that envisioned by Mackinnon for the dBM wave packet associated with a stationary particle, in which the measurements recorded by observers traveling at different velocities are communicated to the common rest frame [Fig.~\ref{Fig:ConceptofDBM}(d)].

When the dipole moves at a velocity $v$ along the $z$-axis with respect to stationary detectors encircling it, each constituent plane-wave undergoes a different Doppler shift in the rest frame of the detectors. The field still comprises plane waves travelling at angles $\varphi$ extending from 0 to $2\pi$, but each plane wave now has a \textit{different} frequency $\omega$. Nevertheless, the new spectral support for the dBM wave packet on the light-cone is related to that for the stationary monochromatic dipole. Indeed, the Lorentz transformation associated with the relative motion between the source and detectors tilts the horizontal iso-frequency spectral plane in Fig.~\ref{Fig:OpticaldBMConcept}(a) by an angle $\theta$ with respect to the $k_{z}$-axis as shown in Fig.~\ref{Fig:OpticaldBMConcept}(b), where $\tan{\theta}\!=\!\beta_{v}$ \cite{Belanger86JOSAA,Longhi04OE,Saari04PRE,Kondakci18PRL}, thus yielding a tilted ellipse whose projection onto the $(k_{x},\tfrac{\omega}{c})$ is:
\begin{equation}\label{Eq:EllipseInFreeSpace}
\frac{k_{x}^{2}}{k_{\mathrm{o}}^{2}}+\frac{(\omega-ck_{1})^{2}}{(\Delta\omega/2)^{2}}=1.
\end{equation}
The spectral projection onto the $(k_{z},\tfrac{\omega}{c})$-plane is now the line $k_{z}\!=\!k_{+}+\tfrac{\omega-\omega_{+}}{\widetilde{v}}\!=\!\tfrac{1}{\beta_{v}}(\tfrac{\omega}{c}-\tfrac{k_{\mathrm{o}}^{2}}{k_{1}})$, where $\widetilde{v}\!=\!c\tan{\theta}\!=\!v$ is the wave-packet group velocity along $z$, $k_{+}\!=\!\tfrac{\omega_{+}}{c}\!=\!k_{\mathrm{o}}\sqrt{\tfrac{1+\beta_{v}}{1-\beta_{v}}}$, and $k_{1}\!=\!k_{\mathrm{o}}/\sqrt{1-\beta_{v}^{2}}$. The spatial and temporal bandwidths are related via $\tfrac{\Delta\omega}{c}\!=\!\beta_{v}\Delta k_{z}$, where $\Delta k_{z}\!=\!2k_{1}$. Each plane wave travels at a different direction in the $(x,z)$-plane in such a way that their \textit{axial} wave numbers $k_{z}$ reproduce the dBM dispersion relationship [compare Fig.~\ref{Fig:ConceptofDBM}(f) to Fig.~\ref{Fig:OpticaldBMConcept}(b)]. By setting the complex spectral amplitudes constant for all frequencies, we obtain the dBM wave packet (with $\widetilde{v}<c$):
\begin{equation}\label{Eq:OpticaldBM}
\psi(x,z;t)\propto e^{i\beta_{v}\Delta k_{z}(z-\widetilde{v}t)}\;\mathrm{sinc}\left(\frac{\Delta k_{z}}{\pi}\sqrt{x^{2}+(z-\widetilde{v}t)^{2}}\right),
\end{equation}

Two parameters uniquely identify the optical dBM wave packet: the group velocity $\widetilde{v}$ (corresponding to the particle velocity) and the wave number $k_{\mathrm{o}}$ (corresponding to the particle mass). Furthermore, the signature of the dBM wave packet in Eq.~\ref{Eq:OpticaldBM} is its circularly symmetric spatio-temporal profile in $(x,t)$-space in any axial plane $z$. In contrast, all other propagation-invariant wave packets that have been observed in free space are X-shaped \cite{Saari97PRL,Porras03PRE2,Turunen10PO,FigueroaBook14,Yessenov22AOP} and are \textit{not} circularly symmetric. Indeed, truncating the spectrum of the optical dBM wave packet obstructs the formation of the circularly symmetric profile and gives rise instead to the more familiar X-shaped counterpart \cite{ZamboniRached2008PRA,Yessenov22AOP}. The O-shaped spatio-temporal profile as indicated by Eq.~\ref{Eq:OpticaldBM} can be observed only when the full bandwidth -- delimited by the light-lines -- is included.

The field in the $(x,z)$-plane recorded by stationary detectors encircling the dipole takes the form shown in Fig.~\ref{Fig:OpticaldBMConcept}(b), as pointed out recently in a thought experiment by Wilczek \cite{WilczekBook}. Despite the conceptual simplicity of this optical scheme for producing dBM wave packets, it nevertheless faces obvious experimental challenges. Encircling an optical dipole moving at a relativistic speed with stationary detectors is far from practical realizability. The more realistic configuration in which the detectors are restricted to a small angular range within the paraxial regime centered on the $z$-axis truncates the recorded field and precludes observing of the O-shaped spatio-temporal profile \cite{ZamboniRached2008PRA,Yessenov22AOP}. For these reasons, it is not expected that the O-shaped dBM wave packet can be observed using spatio-temporally structured optical fields in free space.

%\clearpage
\subsection*{Optical de Broglie-Mackinnon wave packets in a dispersive medium}

\noindent
The necessity of including the entire bandwidth delimited by the intersection of the dBM dispersion relationship with the free-space light-cone [Fig.~\ref{Fig:OpticaldBMConcept}(a-b)] presents insurmountable experimental obstacles. Producing paraxial dBM wave packets necessitates confining the spectrum to a narrow range of values of $k_{z}$ centered at a value $k_{z}\!\sim\!k_{\mathrm{o}}\!>\!0$. Crucially, the linear spatio-temporal spectrum projected onto the $(k_{z},\tfrac{\omega}{c})$-plane must remain delimited at both ends by the light-line, to produce the circularly symmetric spatio-temporal wave-packet profile. Clearly these requirements cannot be met in free space. Nevertheless, this challenge can be tackled by exploiting the unique features of optical-wave propagation in the presence of anomalous GVD. Specifically, the light-cone structure is modified in presence of anomalous GVD so that the curvature of the light-line has the same sign as that of the de~Broglie dispersion relationship [Fig.~\ref{Fig:OpticaldBMConcept}(c)]. In this case, imposing the characteristically linear dBM dispersion relationship produces a spectral support domain on the dispersive light-cone surface that satisfies all the above-listed requirements: (1) $k_{z}\!>\!0$ is maintained throughout the entire span of propagation angles $\varphi(\omega)$; (2) the field simultaneously remains within the paraxial regime; and (3) the spectrum is delimited at both ends by the light-line [Fig.~\ref{Fig:OpticaldBMConcept}(c)], thus yielding a wave packet having a circularly symmetric spatio-temporal profile. The spectral support is in the form of an ellipse at the intersection of the dispersive light-cone with a tilted spectral plane. The center of this ellipse is displaced to a large value of $k_{z}$, and the spectral projection onto the $(k_{z},\tfrac{\omega}{c})$-plane is a line making an angle $\theta$ with the $k_{z}$-axis. The resulting wave packet is propagation-invariant in the dispersive medium and travels at a velocity $\widetilde{v}\!=\!c\tan{\theta}$ independently of the physical parameters of the dispersive medium.

In the anomalous-GVD regime, the wave number is given by $k(\omega)\!=\!n(\omega)\omega/c=k(\omega_{\mathrm{o}}+\Omega)\!\approx\!n_{\mathrm{m}}k_{\mathrm{o}}+\tfrac{\Omega}{\widetilde{v}_{\mathrm{m}}}-\tfrac{1}{2}|k_{2\mathrm{m}}|\Omega^{2}+\cdots$; where $n(\omega)$ is the refractive index, and the following quantities are all evaluated at $\omega\!=\!\omega_{\mathrm{o}}$: $n_{\mathrm{m}}\!=\!n(\omega_{\mathrm{o}})$ is the refractive index, $\widetilde{v}_{\mathrm{m}}\!=\!1\big/\tfrac{dk}{d\omega}\big|_{\omega_{\mathrm{o}}}$ is the group velocity for a plane-wave pulse in the medium, and $k_{2\mathrm{m}}\!=\!\tfrac{d^{2}k}{d\omega^{2}}\big|_{\omega_{\mathrm{o}}}\!=\!-|k_{2\mathrm{m}}|$ is the negative-valued anomalous GVD coefficient \cite{SalehBook07}. The dispersion relationship in the medium $k_{x}^{2}+k_{z}^{2}\!=\!k^{2}$ corresponds geometrically to the surface of the modified dispersive light-cone in Fig.~\ref{Fig:OpticaldBMConcept}(c). Similarly to the free-space scenario, we impose a spectral constraint of the form $k_{z}\!=\!n_{\mathrm{m}}k_{\mathrm{o}}+\tfrac{\Omega}{\widetilde{v}}\!=\!\tfrac{1}{\beta_{v}}\left\{\tfrac{\omega}{c}-k_{\mathrm{o}}(1-n_{\mathrm{m}}\beta_{v})\right\}$ in the medium, where $\Omega\!=\!\omega-\omega_{\mathrm{o}}$ and $\widetilde{v}\!=\!c\tan{\theta}$ is the group velocity of the wave packet [Fig.~\ref{Fig:OpticaldBMConcept}(c)]. The wave-packet spectrum as defined by this constraint is delimited by the light-line at its two ends, both located however in the range $k_{z}\!>\!0$, in contrast to the previous scenarios depicted in Fig.~\ref{Fig:ConceptofDBM}(e,f) and Fig.~\ref{Fig:OpticaldBMConcept}(a,b); see Methods.  

The spectral projections onto the $(k_{x},\tfrac{\omega}{c})$ and $(k_{x},k_{z})$ planes of the spectral support on the dispersive light-cone are ellipses (Methods):
\begin{equation}\label{Eq:DispersiveEllipseOmegaKx}
\frac{k_{x}^{2}}{k_{x,\mathrm{max}}^{2}}+\frac{(\omega-\omega_{\mathrm{c}})^{2}}{(\Delta\omega/2)^{2}}=1,\quad \frac{k_{x}^{2}}{k_{x,\mathrm{max}}^{2}}+\frac{(k_{z}-k_{\mathrm{c}})^{2}}{(\Delta k_{z}/2)^{2}}=1,
\end{equation}
where the temporal bandwidth is $\tfrac{\Delta\omega}{c}\!=\!2\tfrac{k_{\mathrm{o}}}{\sigma_{\mathrm{m}}}\tfrac{1-\beta_{v}'}{\beta_{v}^{2}}\!=\!\beta_{v}\Delta k_{z}$, $\sigma_{\mathrm{m}}\!=\!c\omega_{\mathrm{o}}|k_{2\mathrm{m}}|$ is a dimensionless dispersion coefficient, $\beta_{v}'\!=\!\tfrac{\widetilde{v}}{\widetilde{v}_{\mathrm{m}}}$, $k_{x,\mathrm{max}}\!=\!\tfrac{1}{2}\tfrac{\Delta\omega}{c}\sqrt{n_{\mathrm{m}}\sigma_{\mathrm{m}}}$, $\omega_{\mathrm{c}}\!=\!\omega_{\mathrm{o}}-\Delta\omega/2$, $k_{\mathrm{c}}\!=\!n_{\mathrm{m}}k_{\mathrm{o}}-\tfrac{\Delta k_{z}}{2}$, $k_{x,\mathrm{max}}\!\ll\!n_{\mathrm{m}}k_{\mathrm{o}}$, $\Delta k_{z}\!\ll\!n_{\mathrm{m}}k_{\mathrm{o}}$, and $\Delta\omega\!\ll\!\omega_{\mathrm{o}}$. It is crucial to recognize that the ellipse projected onto the $(k_{x},k_{z})$-plane does \textit{not} enclose the origin $(k_{x},k_{z})\!=\!(0,0)$, but is rather displaced to a central value $k_{\mathrm{c}}\!\gg\!\Delta k_{z}$. Therefore, the optical field comprises plane-wave components that propagate only in the forward direction within a small angular range centered on the $z$-axis, and the field thus remains within the paraxial domain. Nevertheless, because the spectrum is delineated at both ends by the curved dispersive light-line, the resulting spatio-temporal profile is circularly symmetric in any axial plane $z$. This wave packet in the dispersive medium thus satisfies all the above-listed desiderata for an optical dBM wave packet, but can be readily synthesized and observed in contrast to its free-space counterparts. One difficulty, however, arises from the form of $\varphi(\omega)$ in the dispersive medium, which differs fundamentally from that in free space. Each frequency $\omega$ in a free-space optical dBM wave packet is associated with two propagation angles $\pm\varphi(\omega)$. However, each propagation angle $\varphi$ is associated with a single frequency, so that $|\phi(\omega)|$ is one-to-one. In the optical dBM wave packet in the dispersive medium, each $\omega$ is still associated with two propagation angles $\pm\varphi(\omega)$; but $\varphi(\omega)$ is now two-to-one, so that $\varphi(\omega)$ is folded back on itself [Fig.~\ref{Fig:OpticaldBMConcept}(c)]. To synthesize such a field configuration, a synthesizer capable of sculpting $\varphi(\omega)$ almost arbitrarily is required.

%\clearpage
\subsection*{Experimental confirmation}

\noindent\textbf{Setup.} To construct the optical dBM wave packet in free space from a generic pulsed beam in which the spatial and temporal degrees-of-freedom are uncoupled, we introduce angular dispersion by assigning to each wavelength $\lambda$ a particular pair of angles $\pm\varphi(\lambda)$, thereby coupling the spatial and temporal degrees-of-freedom. We carry out this task using a universal angular-dispersion synthesizer \cite{Hall21OE2}, in which a spatial light modulator (SLM) deflects each wavelength from a spectrally resolved laser pulse at prescribed angles, as illustrated in Fig.~\ref{Fig:Setup} (Methods). Because each wavelength $\lambda$ is deflected at $\varphi(\lambda)$ independently of all other wavelengths, $\varphi(\lambda)$ need \textit{not} be one-to-one. Indeed, it can readily be a two-to-one mapping as required for paraxial optical dBM wave packets. The dBM wave packet is formed once all the wavelengths are recombined by a grating to reconstitute the pulsed field. The spatio-temporal spectrum of the synthesized wave packet is acquired by operating on the spectrally resolved field with a spatial Fourier transform and recording the intensity with a CCD camera. This measurement yields the spatio-temporal spectrum projected onto the $(k_{x},\lambda)$-plane, from which we can obtain the spectral projection onto the $(k_{z},\lambda)$-plane. The spatio-temporal envelope $I(x;\tau)$ of the intensity profile at a fixed axial plane $z$ is reconstructed in the frame travelling at $\widetilde{v}$ ($\tau\!=\!t-z/\widetilde{v}$) via linear interferometry exploiting the procedure developed in Refs.~\cite{Kondakci19NC,Yessenov19OE,Bhaduri20NatPhot} (Methods). The dispersive medium exploited in our measurements is formed of a pair of chirped Bragg mirrors providing an anomalous GVD coefficient of $k_{2\mathrm{m}}\!\approx\!-500$~fs$^2$/mm and $\widetilde{v}_{\mathrm{m}}\!\approx\!c$ (Methods).

\begin{figure*}[t!]
\centering
\includegraphics[width=8.6cm]{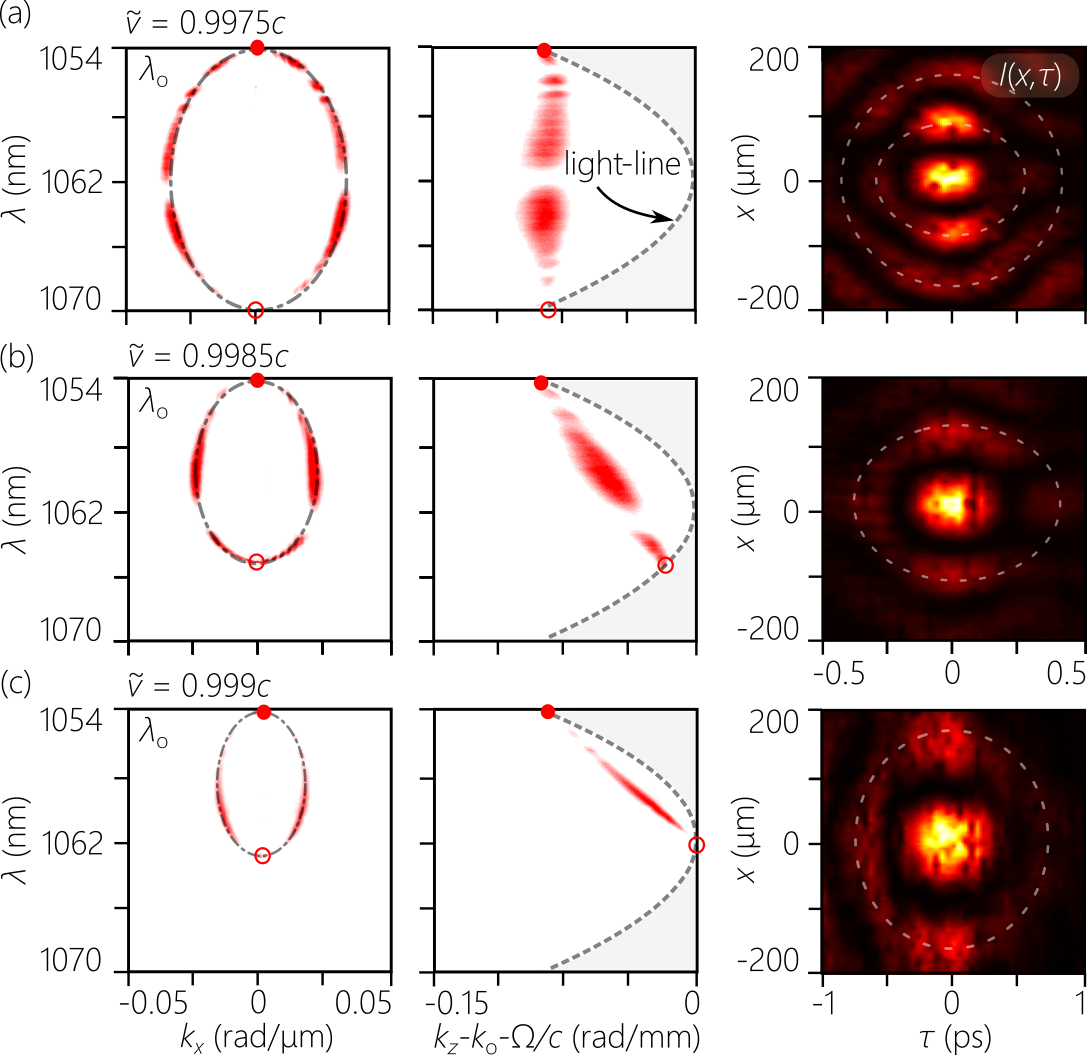}
\caption{\textbf{Observation of optical dBM wave packets in presence of anomalous GVD and tuning their group velocity.} In the first column we plot the measured spatio-temporal spectrum projected onto the $(k_{x},\lambda)$-plane; the dotted curve is the theoretical expectation based on Eq.~\ref{Eq:DispersiveEllipseOmegaKx}. In the second column we plot the spectral projection onto the $(k_{z},\lambda)$-plane using the data from the first column; the dotted curve is the dispersive light-line. In  the third column we plot the spatio-temporal intensity profile $I(x;\tau)$ acquired in a moving reference frame traveling at the group velocity of the wave packets; the dotted circles are guides for the eye. (a) Measurements for an optical dBM wave packet having a group velocity $\widetilde{v}\!=\!0.9975c$; (b) $\widetilde{v}\!=\!0.9985c$; and (c) $\widetilde{v}\!=\!0.999c$. The dimensionaless dispersion coefficient is $\sigma_{\mathrm{m}}\!=\!c\omega_{\mathrm{o}}k_{2\mathrm{m}}=0.3$, and the measurements are carried out at $z\!=\!30$~mm.}
\label{Fig:MeasurementsChangingV}
\end{figure*}

\noindent\textbf{Measurement results.} We first verify the unique signature of dBM wave packets in presence of anomalous GVD; namely, the O-shaped spatio-temporal intensity profile at any axial plane after inculcating into the field the dBM dispersion relationship. In Fig.~\ref{Fig:MeasurementsChangingV} we verify three sought-after features: (1) The closed elliptical spatio-temporal spectrum projected onto the $(k_{x},\lambda)$-plane; (2) the linear spectral projection onto the $(k_{z},\lambda)$-plane, indicating non-dispersive propagation in the dispersive medium; and (3) the circularly symmetric spatio-temporal intensity profile $I(x;\tau)$ reconstructed at a fixed axial plane ($z\!=\!30$~mm). In Fig.~\ref{Fig:MeasurementsChangingV}(a) we plot the measurements for an optical dBM wave packet having a group velocity $\widetilde{v}\!=\!0.9975c$. The temporal bandwidth is constrained to a maximum value of $\Delta\lambda\!\approx\!16$~nm, and the associated spatial bandwidth $\Delta k_{x}\!\approx\!0.03$~rad/$\mu$m, thus resulting in a pulsewidth $\Delta T\!\approx\!200$~fs at $x\!=\!0$, and a spatial profile width $\Delta x\!\approx\!38$~$\mu$m at $\tau\!=\!0$. The spectral projection onto the $(k_{z},\lambda)$-plane is delimited at both ends by the curved light-line of the dispersive medium. In other words, a larger bandwidth is \textit{in}compatible at this group velocity with propagation invariance in the dispersive medium. Further increase in the bandwidth extends the spectral projection \textit{below} the dispersive light-line, which contributes to only evanescent field components. The measured spatio-temporal profile $I(x;\tau)$ therefore has the smallest dimensions in space and time for a circularly symmetric dBM wave packet compatible with the selected group velocity in the medium.

To the best of our knowledge, this is the first observation of an O-shaped spatio-temporal intensity profile for a dispersion-free wave packet in a linear dispersive medium. Previous realizations of dispersion-free ST wave packets in dispersive media (whether in the normal- or anomalous-GVD regimes) revealed X-shaped spatio-temporal profiles \cite{Hall22LPR} similar to those observed in free space \cite{Saari97PRL,Kondakci17NP,Kondakci19NC} or in non-dispersive dielectrics \cite{Bhaduri20NatPhot}. In these experiments, however, the wave packets were \textit{not} delimited spectrally by the dispersive-medium light-line, which is the prerequisite for the realization of O-shaped optical dBM wave packets.

\begin{figure*}[t!]
\centering
\includegraphics[width=8.6cm]{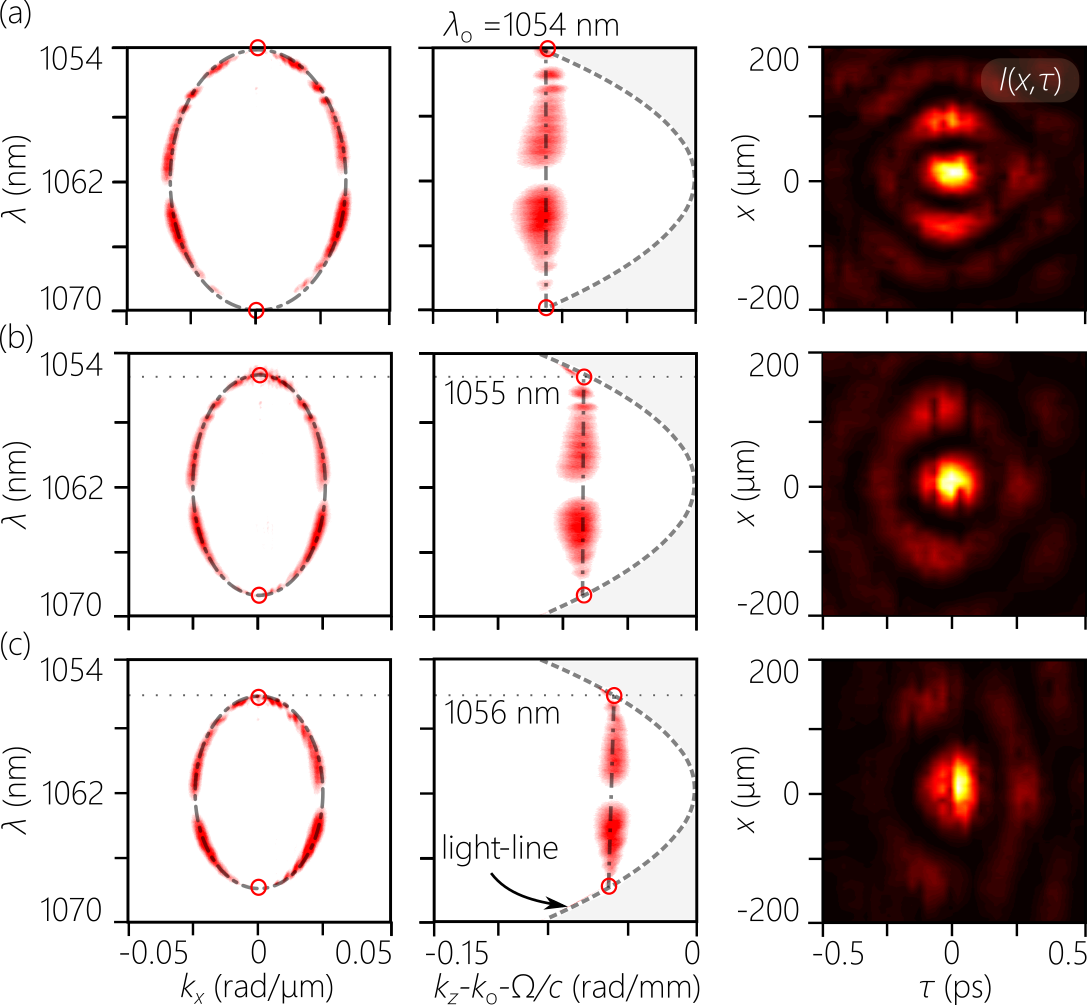}
\caption{\textbf{Tuning the equivalent rest-mass of an optical dBM wave packets.} The columns correspond to those in Fig.~\ref{Fig:MeasurementsChangingV}. The dimensionless dispersion coefficient is $\sigma_{\mathrm{m}}\!=\!c\omega_\mathrm{o}k_{2\mathrm{m}}=0.3$, the measurements are all acquired at $z\!=\!30$~mm, and the group velocity is held fixed at $\widetilde{v}=0.9975c$. (a) The short-wavelength intersection with the dispersive light-line is $\lambda_{\mathrm{o}}\!=\!1054$~nm; (b) $\lambda_{\mathrm{o}}\!=\!1055$~nm; and (c) $\lambda_{\mathrm{o}}\!=\!1056$~nm.}
\label{Fig:MeasurementsFixedV}
\end{figure*}

As mentioned earlier, two parameters characterize a dBM wave packet: the velocity $v$ and the rest mass $m_{\mathrm{o}}$. The corresponding variables associated with the optical dBM wave packet are $\widetilde{v}$ and $\lambda_{\mathrm{o}}$, which can both be readily tuned in our experimental arrangement by changing the functional dependence of $\varphi$ on $\lambda$. In this way we can vary the first parameter; namely, the group velocity $\widetilde{v}$. Increasing the group velocity from $\widetilde{v}\!=\!0.9975c$ [Fig.~\ref{Fig:MeasurementsChangingV}(a)] to $\widetilde{v}\!=\!0.9985c$ [Fig.~\ref{Fig:MeasurementsChangingV}(b)] and then to $\widetilde{v}\!=\!0.999c$ [Fig.~\ref{Fig:MeasurementsChangingV}(c)] reduces the maximum exploitable temporal bandwidth from $\Delta\lambda\!\approx\!16$~nm to $\Delta\lambda\!\approx\!8$~nm and $\Delta\lambda\!\approx\!6$~nm, respectively, while retaining the closed elliptic spectral projection onto the $(k_{x},\lambda)$-plane, the linear spectral projection onto the $(k_{z},\lambda)$-plane, and the associated O-shaped spatio-temporal profile $I(x;\tau)$. The corresponding spatial bandwidths drop to $\Delta k_{x}\!\approx\!0.023$~rad/$\mu$m and $\Delta k_{x}\!\approx\!0.017$~rad/$\mu$m, respectively. In all three dBM wave packets in Fig.~\ref{Fig:MeasurementsChangingV}, we retain a fixed intersection with the dispersive light-line at $\lambda_{\mathrm{o}}\!\approx\!1054$~nm (corresponding to a fixed particle mass), such that reducing $\widetilde{v}$ decreases the wavelength of the second intersection point. The second parameter, the wavelength $\lambda_{\mathrm{o}}$ corresponding to particle rest mass $m_{\mathrm{o}}$ for de~Broglie phase waves, can also be readily tuned [Fig.~\ref{Fig:MeasurementsFixedV}]. Here, the maximum exploitable bandwidth changes as a result of shifting the value of $\lambda_{\mathrm{o}}$ from $\lambda_{\mathrm{o}}\!=\!1054$~nm [Fig.~\ref{Fig:MeasurementsFixedV}(a)] where $\Delta\lambda\!=\!16$~nm, to $\lambda_{\mathrm{o}}\!=\!1055$~nm [Fig.~\ref{Fig:MeasurementsFixedV}(b)] where $\Delta\lambda\!=\!14$~nm, and then to $\lambda_{\mathrm{o}}\!=\!1056$~nm [Fig.~\ref{Fig:MeasurementsFixedV}(c)] where $\Delta\lambda\!=\!12$~nm. Once again, both the spatial and temporal widths of the circularly symmetric O-shaped profile in the $(x,t)$-domain change accordingly.

\begin{figure*}[t!]
\centering
\includegraphics[width=17cm]{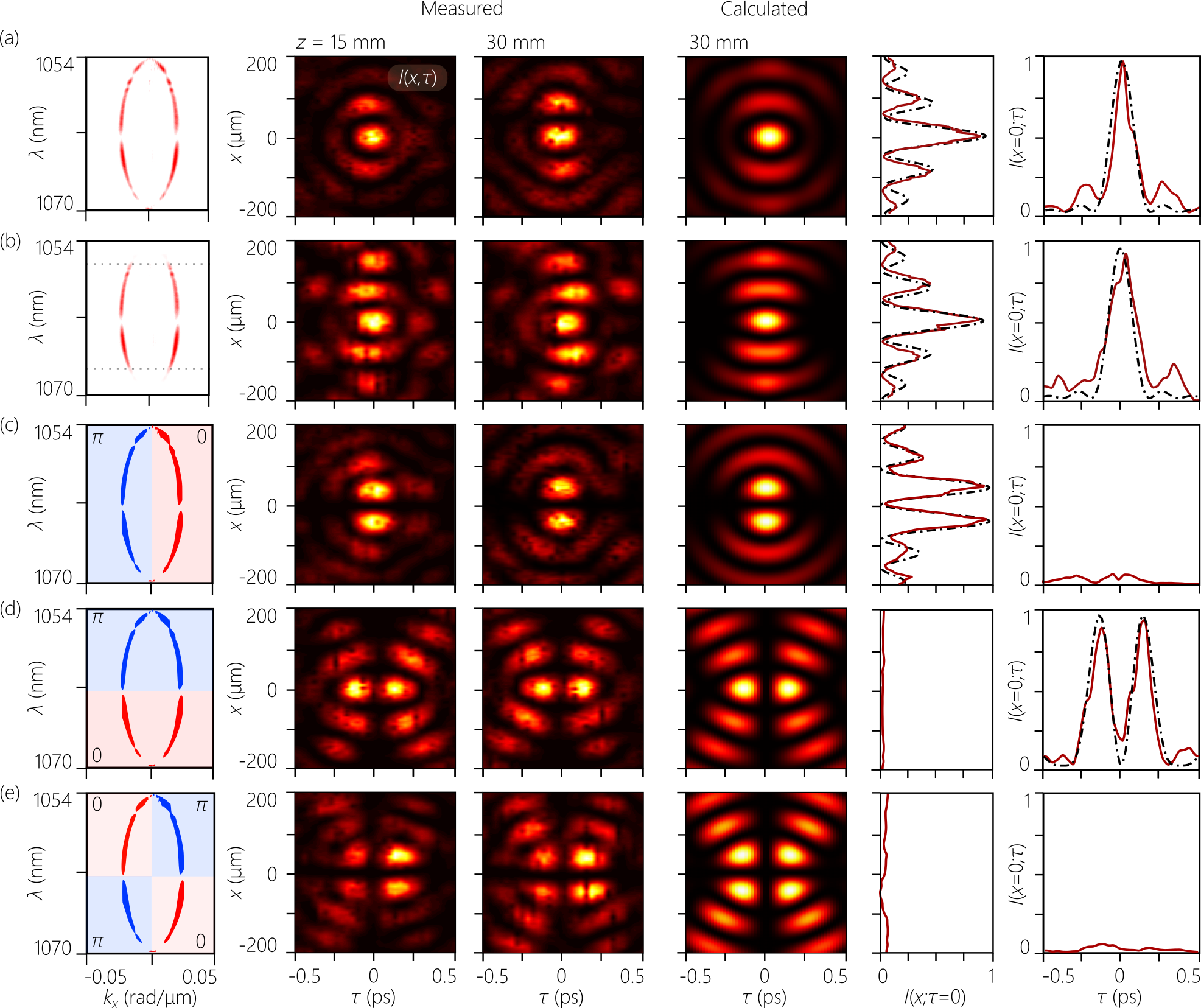}
\caption{\textbf{Changing the spatio-temporal structure of optical dBM wave packets.} The columns from left to right are: the measured spatio-temporal spectrum projected onto the $(k_{x},\lambda)$-plane, along with the spectral phase; the measured spatio-temporal intensity profile $I(x,z;\tau)$ at $z\!=\!15$~mm; the measured spatio-temporal intensity profile $I(x,z;\tau)$ at $z\!=\!30$~mm; the calculated spatio-temporal intensity profile $I(x,z;\tau)$ at fixed $z$; $I(x;\tau=0)$ at $z\!=\!30$~mm; and $I(x=0;\tau)$ at $z\!=\!30$~mm. The dimensionless dispersion coefficient is  Measurement of an O shaped spectrum with $v_a=0.9975c$ and $\sigma_{\mathrm{m}}\!=\!c\omega_{\mathrm{o}}k_{2\mathrm{m}}=0.3$, and the group velocity is held fixed at $\widetilde{v}\!=\!0.9975c$, corresponding to Fig.~\ref{Fig:MeasurementsChangingV}(a). (a) The spectrum has uniform phase; (b) the spectrum has uniform phase but its amplitude is truncated along $\lambda$; (c) a $\pi$-phase step is introduced along $k_{x}$; (d) a $\pi$-phase step is introduced along $\lambda$; and (e) the spectral phase is alternated between $0$ and $\pi$ in  the four quadrants of the $(k_{x},\lambda)$-plane.}
\label{Fig:ChangingProfile}
\end{figure*}

The Airy wave packet, as mentioned earlier, is the \textit{unique} non-dispersive solution to Schr{\"o}dinger's equation -- no other waveform will do \cite{Unnikrishnan96AJP}. Although Mackinnon obtained a particular `sinc'-function-shaped wave packet \cite{Mackinnon78FP}, this waveform is \textit{not} unique. Indeed, the sinc-function results from combining all the de~Broglie phase waves with equal weights. However, dBM wave packets can take on in principle arbitrary waveforms by associating different magnitudes or phases with the plane-wave components constituting it. We confirm in Fig.~\ref{Fig:ChangingProfile} that the spatio-temporal profile $I(x;\tau)$ of optical dBM wave packets can be modified while remaining propagation invariant in the dispersive medium. First, setting the complex spectral amplitudes equal along the elliptical spectral support, we obtain propagation-invariant circularly symmetric wave packets in the dispersive medium [Fig.~\ref{Fig:ChangingProfile}(a)]. Truncating the ellipse and eliminating the plane wave components in the vicinity of $k_{x}\!=\!0$ disrupts the formation of the full circular profile, but the wave packet nevertheless propagates invariantly [Fig.~\ref{Fig:ChangingProfile}(b)]. By introducing a $\pi$-step in the spectral phase along $k_{x}$, a spatial null is formed along $x\!=\!0$ in the profile of the dBM wave packet [Fig.~\ref{Fig:ChangingProfile}(c)], whereas introducing the $\pi$-phase-step along $\lambda$ produces a temporal null along $\tau\!=\!0$ [Fig.~\ref{Fig:ChangingProfile}(d)]. Finally, alternating the phases between 0 and $\pi$ in the four quadrants of the spatio-temporal spectral plane $(k_{x},\lambda)$ produces spatial and temporal nulls along $x\!=\!0$ and $\tau\!=\!0$, respectively [Fig.~\ref{Fig:ChangingProfile}(e)]. Despite such variations in their spatio-temporal profiles, all these optical dBM wave packets propagate invariantly in the dispersive medium.

%\clearpage
\section*{Discussion}

\noindent

The rapidly evolving versatile techniques for synthesizing optical fields \cite{Forbes21NP,Yessenov22AOP} played a critical role in the realization of dBM wave packets as demonstrated here. This has helped confirm the theoretical proposal made by Mackinnon almost 45 years ago for constructing a non-dispersive wave packet from dispersive de~Broglie phase waves \cite{Mackinnon78FP}. Furthermore, the experimental procedure implemented here points to a general synthesis strategy that extends beyond the particular scenario of dBM wave packets. The overarching theme is that novel dispersion relationships for the axial propagation of a wave packet can be imposed by first adding another dimension to the space, and then exploiting the new dimension to tailor the dispersion relationship before spectral projection back onto the original reduced-dimensionality space.

In the scenario studied here, we start with a $(1+1)$D physical wave in which an axial dispersion relationship $\omega(k_{z})$ is enforced by the dynamics of the wave equation. Increasing the dimensionality of the space from $(1+1)$D to $(2+1)$D by including a transverse coordinate $x$ yields a new dispersion relationship $\omega(k_{x},k_{z})$. In free space, optical wave packets are subject to the constraint $\omega\!=\!ck_{z}$ in $(1+1)$D and $\omega(k_{x},k_{z})\!=\!c\sqrt{k_{x}^{2}+k_{z}^{2}}$ in $(2+1)$D. Now, by judiciously associating each transverse wave number $k_{x}$ with a particular axial wave number $k_{z}$, a reduced-dimensional axial dispersion relationship $\omega_{\mathrm{red.}}(k_{z})$ is obtained: $\omega(k_{x},k_{z})\!=\!\omega(k_{x}(k_{z}),k_{z})\!\mapsto\!\omega_{\mathrm{red.}}(k_{z})$, which can be engineered almost arbitrarily. In the experiment reported here, we employed this strategy to produce a linear dispersion relationship $\omega(k_{z})\!=\!(k_{z}-k_{\mathrm{o}})\widetilde{v}$ projected onto the $(k_{z},\tfrac{\omega}{c})$-plane that deviates away from the light-line $\omega\!=\!ck_{z}$. In presence of anomalous GVD, such a spatio-temporal spectrum is delimited at both ends by the curved light-line in the dispersive medium -- thereby yielding the circular symmetric spatio-temporal profile characteristic of dBM wave packets. Here, the transverse wave number $k_{x}$ played the role of the observer velocity $u$ in the physical configuration envisioned by Mackinnon [Fig.~\ref{Fig:ConceptofDBM}(d)]. However, one may envision a variety of other scenarios that can be facilitated by this general strategy. For example, besides tuning the group velocity in free space, linear dispersive media, or nonlinear optical materials and structures, one may produce accelerating wave packets \cite{Yessenov20PRLaccel,Hall22OLaccel,Li20SR} whose group velocity changes with propagation in such media. These features have been recently predicted to produce a host of new phenomena related to two photon-emission \cite{Sloan22NatPhys} and relativistic optics \cite{Bliokh12PRA,Caloz20IEEE}.

Intriguingly, the strategy employed here is not constrained to optical waves. Indeed, our approach to spatio-temporal structuring of the field is agnostic with respect to the physical substrate, and can be implemented in principle with acoustic waves, microwaves, surface plasmon polaritons \cite{Schepler20ACSPhot}, electron beams, neutron beams, or other massive particles. In all cases, an added spatial dimension can be exploited to override the intrinsic dispersion relationship of the particular wave phenomenon, thus producing novel propagation dynamics.

The dimensionality of the $(2+1)$D dBM wave packets synthesized here can be further extended to the full-dimensional $(3+1)$D space of $(x,y,z;t)$ by including the second transverse coordinate $y$. This can now be achieved in light of very recent progress in producing so-called 3D ST wave packets that are localized in all dimensions of $(3+1)$D space \cite{Guo21Light,Pang22OE,Yessenov22NC}. Combining this new synthesis methodology with the procedure outlined here for producing dBM wave packets in the anomalous-GVD regime will yield spherically symmetric propagation-invariant pulsed field structures. Such field configurations provide a platform for exploring proposed topological structures associated with polarization (spin texture) \cite{Guo21Light} \textit{without} resorting to stereo-projection onto a 2D plane. Moreover, such spherically symmetric optical dBM wave packets are compatible with coupling to optical fibers and waveguides, thus enabling new opportunities in optical communications, optical signal processing, and nonlinear and quantum optics.

Finally, the ideal spectral constraint underlying optical dBM wave packets implies an exact association between the spatial and temporal frequencies. Such idealized wave packets consequently have infinite energy \cite{Sezginer85JAP}. In any realistic system, however, a spectral uncertainty is inevitably introduced into this association, resulting in a finite-energy wave packet traveling for a finite distance over which it is approximately invariant \cite{Kondakci19OLClassic}. In our experiments, this spectral uncertainty arises from the finite spectral resolution of the diffraction grating employed (Fig.~\ref{Fig:Setup}), which is estimated to be $\approx\!16$~pm, corresponding to a propagation distance of $\approx\!32$~m at a spectral tilt angle $\theta=44.99^{\circ}$ \cite{Yessenov19OE}.

%\clearpage
\bibliography{diffraction}

\vspace{4mm}
\noindent
\textbf{Data availability statement}\\
The data that support the plots within this paper and other findings of this study are available from the corresponding author upon reasonable request.

\vspace{2mm}
\noindent
\textbf{Acknowledgments}\\
We thank M. Yessenov, K. L. Schepler, D. N. Christidoulides and A. Dogariu for helpful discussions. This work was supported by the U.S. Office of Naval Research (ONR) under contracts N00014-17-1-2458 and N00014-20-1-2789.
\vspace{2mm}\\
\noindent
\textbf{Author contributions}\\
\noindent

\noindent
Correspondence and requests for materials should be addressed to A.F.A.\\(email: raddy@creol.ucf.edu)

\vspace{2mm}
\noindent
\textbf{Competing interests:} The authors declare no competing interests.

\clearpage

\section*{Methods}

\noindent\textbf{de~Broglie phase waves.} At rest, the frequency of the internal `clock' and that of the infinite-wavelength phase wave are the de Broglie frequency $\omega_{\mathrm{o}}$. When the particle moves at a velocity $v$, the observed frequency of the internal `clock' in the rest frame is reduced to $\omega\!=\!\omega_{\mathrm{o}}\sqrt{1-\beta_{v}^{2}}$ whereas that of the phase wave is increased to $\omega\!=\!\omega_{\mathrm{o}}/\sqrt{1-\beta_{v}^{2}}$, where $\beta_{v}\!=\!v/c$ []. To obtain The phase velocity $v_{\mathrm{phh}}$ of the phase wave, de~Broglie proposed a `theory of phase harmony', which requires that the internal clock and the phase wave remain in phase for all $t$ and at any $v$ []. The phase of the moving clock in the rest frame is $\phi\!=\!\omega_{\mathrm{o}}t\sqrt{1-\beta_{v}^{2}}$, and that of the phase wave is $\phi\!=\!\omega_{\mathrm{o}}(t-\tfrac{z}{v_{\mathrm{ph}}})/\sqrt{1-\beta_{v}^{2}}$. At time $t$, the particle has covered a distance $z\!=\!vt$, and equating the phases yields $v_{\mathrm{ph}}\!=\!\tfrac{c^{2}}{v}\!>\!c$. Alternatively, the Lorentz transformation of the proper time is $t'\!=\!(t-\tfrac{v}{c^{2}}z)/\sqrt{1-\beta_{v}^{2}}$, $\omega_{\mathrm{o}}t'\!=\!\omega t-kz\!=\!\omega t-\tfrac{\omega}{v_{\mathrm{ph}}}z$, from which we again have $v_{\mathrm{ph}}\!=\!\tfrac{c^{2}}{v}$.

\noindent\textbf{Conventional de~Broglie wave packets.} The equations $\omega\!=\!\omega_{\mathrm{o}}/\sqrt{1-\beta_{v}^{2}}$ and $k\!=\!\tfrac{\omega}{v_{\mathrm{ph}}}\!=\!\tfrac{\omega}{c}\beta$ for the phase wave yield the dispersion relationship $k\!=\!\tfrac{1}{c}\sqrt{\omega^{2}-\omega_{\mathrm{o}}^{2}}$. A de~Broglie wave packet of finite bandwidth $\Delta\omega$ centered at $\omega\!=\!\omega_{\mathrm{c}}$ requires including an uncertainty $\Delta v$ in the particle velocity centered on the speed $v$ (which corresponds to $\omega_{\mathrm{c}}$). The group velocity of the de~Broglie wave packet is thus $\widetilde{v}\!=\!1\big/\tfrac{dk}{d\omega}\big|_{\omega_{\mathrm{c}}}\!=\!c\sqrt{1-(\tfrac{\omega_{\mathrm{o}}}{\omega_{\mathrm{c}}})^{2}}\!=\!c\beta_{v}\!=\!v$.

\begin{figure*}[t!]
\centering
\includegraphics[width=16cm]{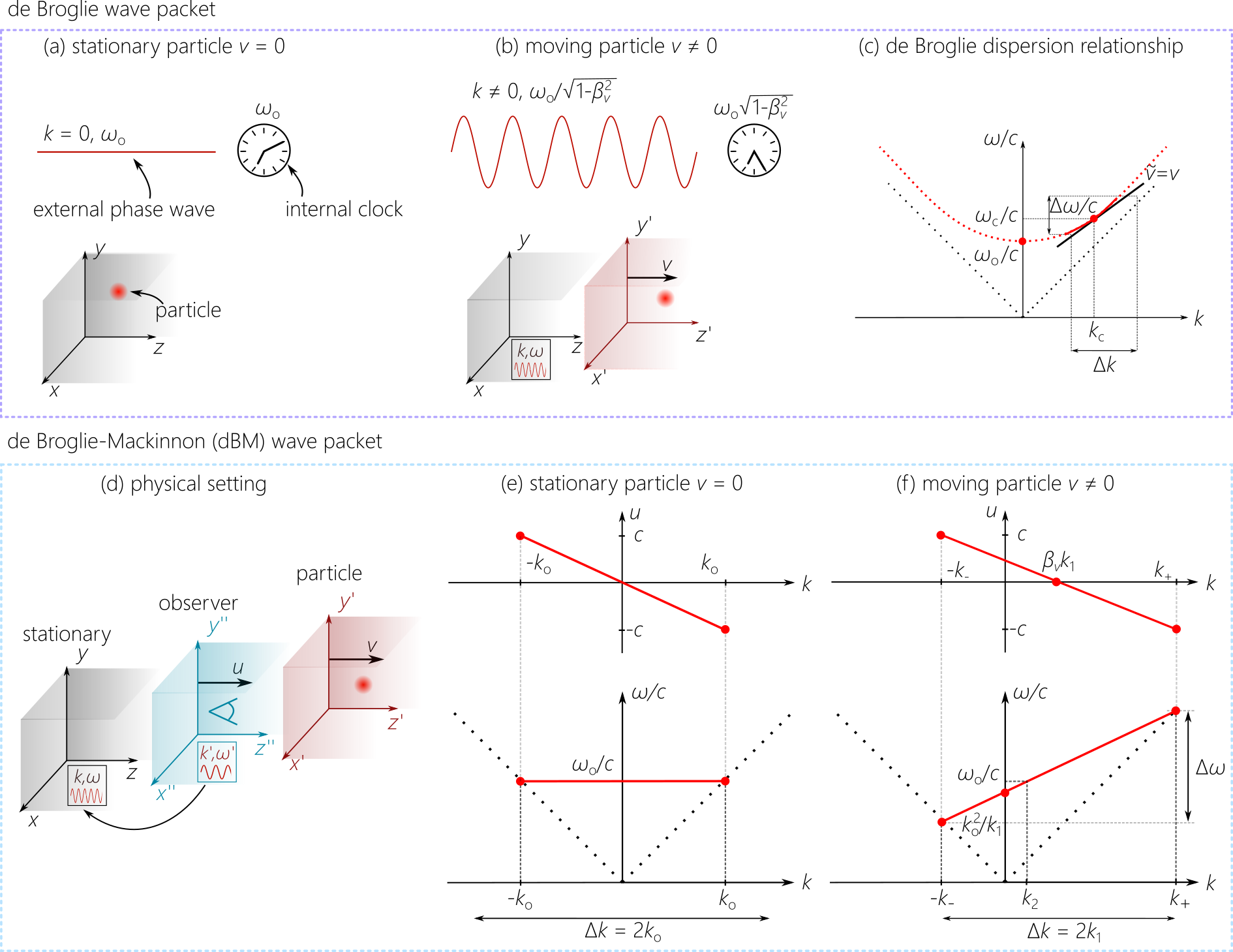}
\caption{\textbf{de~Broglie phase waves and wave packets, and de~Broglie-Mackinnon (dBM) wave packets.} (a) In the rest frame of a particle, the internal `clock' and the external `phase wave' theorized by de~Broglie both have the same frequency $\omega_{\mathrm{o}}$. (b) When the particle moves at a velocity $v$ along $z$, the frequency of the internal clock in the rest frame decreases to $\omega\!=\!\omega_{\mathrm{o}}\sqrt{1-\beta_{v}^{2}}$, whereas that of the phase wave increases to $\omega\!=\!\omega_{\mathrm{o}}/\sqrt{1-\beta_{v}^{2}}$. (c) The dispersion relationship for de~Broglie phase waves $\omega^{2}\!=\!\omega_{\mathrm{o}}^{2}+c^{2}k^{2}$ plotted in $(k,\tfrac{\omega}{c})$-space. The group velocity evaluated at $\omega\!=\!\omega_{\mathrm{c}}$ is $\widetilde{v}\!=\!v$. Constructing a localized de~Broglie wave packet necessitates introducing an \textit{ad hoc} uncertainty in the particle velocity. (d) The physical setting for a dBM wave packet. The particle travels at $v$ and the observer at $u$ (both along the $z$-axis) with respect to a common rest frame. (e) The dispersion relationship for a dBM wave packet in $(k,\tfrac{\omega}{c})$-space (lower panel) corresponding to a stationary particle $v\!=\!0$, delimited by the light-lines $|k|\!=\!\tfrac{\omega}{c}$. The observer velocity $u$ (upper panel) is an internal parameter swept from $-c$ to $c$ to produce the dBM dispersion relationship. (g) Same as (f) for $v\!\neq\!0$; here $k_{+}\!=\!k_{\mathrm{o}}\sqrt{\tfrac{1+\beta_{v}}{1-\beta_{v}}}$, $k_{-}\!=\!k_{\mathrm{o}}\sqrt{\tfrac{1-\beta_{v}}{1+\beta_{v}}}$, $k_{1}\!=\!\tfrac{k_{\mathrm{o}}}{\sqrt{1-\beta_{v}^{2}}}$, and $k_{2}\!=\!\tfrac{k_{\mathrm{o}}}{\beta_{v}}(1-\tfrac{k_{\mathrm{o}}}{k_{1}})$; see Methods}
\label{Fig:ConceptofDBM}
\end{figure*}

\noindent\textbf{Formulation of de~Broglie-Mackinnon (dBM) wave packets.} Consider the configuration depicted in Fig.~\ref{Fig:ConceptofDBM}(d), where the particle moves at a velocity $v$ and an observer moves at a velocity $u$, both with respect to a selected rest frame. Here the relative velocity of the particle with respect to the observer is $\xi$, where $\beta_{\xi}\!=\!\tfrac{\xi}{c}\!=\!\tfrac{\beta_{v}-\beta_{u}}{1-\beta_{v}\beta_{u}}$, $\beta_{v}\!=\!\tfrac{v}{c}$, and $\beta_{u}\!=\!\tfrac{u}{c}$. According to this observer, the frequency and wave number are $\omega'\!=\!\omega_{\mathrm{o}}\big/\sqrt{1-\beta_{\xi}^{2}}$ and $k'\!=\!k_{\mathrm{o}}\beta_{\xi}\big/\sqrt{1-\beta_{\xi}^{2}}$, respectively. The crucial step proposed by Mackinnon is that \textit{all} potential observers with velocities $u$ ranging from $-c$ to $c$ report their observations of $\omega'$ and $k'$ to the common rest frame, where the wave packet is constructed epistemologically after accounting for Lorentz contraction and time dilation. Following this prescription, it is straightforward to show that
\begin{equation}
\omega=\omega_{\mathrm{o}}\frac{1-\beta_{v}\beta_{u}}{\sqrt{1-\beta_{v}^{2}}},\quad k=k_{\mathrm{o}}\frac{\beta_{v}-\beta_{u}}{\sqrt{1-\beta_{v}^{2}}}.
\end{equation}
Because $v$ is a fixed velocity whereas $u$ extends from $-c$ to $c$, a linear dispersion relationship between $\omega$ and $k$ is established [Fig.~\ref{Fig:ConceptofDBM}(f)],
\begin{equation}
k=\frac{1}{\beta_{v}}\left(\tfrac{\omega}{c}-k_{\mathrm{o}}\sqrt{1-\beta_{v}^{2}}\right).
\end{equation}
This line intersects with the light-line $k\!=\!\tfrac{\omega}{c}$ at $k\!=\!k_{+}\!=\!k_{\mathrm{o}}\sqrt{\tfrac{1+\beta_{v}}{1-\beta_{v}}}$ (when $u\!=\!-c$), and with the light-line $k\!=\!-\tfrac{\omega}{c}$ at $k\!=\!-k_{-}\!=\!-k_{\mathrm{o}}\sqrt{\tfrac{1-\beta_{v}}{1+\beta_{v}}}$ (when $u\!=\!c$), where $k_{+}\!=\!\tfrac{\omega_{+}}{c}\!>\!k_{\mathrm{o}}$ and $k_{-}\!=\!\tfrac{\omega_{-}}{c}\!<\!k_{\mathrm{o}}$ [Fig.~\ref{Fig:ConceptofDBM}(f)]. When $u\!=\!0$, the associated de~Broglie phase wave has $\tfrac{\omega}{c}\!=\!k_{1}$ and $k\!=\!\beta_{v}k_{1}$, where $k_{1}\!=\!\tfrac{k_{\mathrm{o}}}{\sqrt{1-\beta_{v}^{2}}}\!>\!k_{\mathrm{o}}$. Another phase wave of interest is the one that retains the stationary frequency $\omega\!=\!\omega_{\mathrm{o}}$, which occurs when $\beta_{u}\!=\!\tfrac{1}{\beta_{v}}(1-\sqrt{1-\beta_{v}^{2}})\!=\!\tfrac{1}{\beta_{v}}(1-\tfrac{k_{\mathrm{o}}}{k_{1}})$, and is associated with $k\!=\!k_{2}\!=\!k_{\mathrm{o}}\beta_{u}$. Finally, the linear dispersion relationship has $k\!=\!0$ when $u\!=\!v$ and thus $\tfrac{\omega}{c}\!=\!k_{\mathrm{o}}\sqrt{1-\beta_{v}^{2}}\!=\!\tfrac{k_{\mathrm{o}}^{2}}{k_{1}}\!<\!k_{\mathrm{o}}$. Throughout, setting $v\!=\!0$ ($\beta_{v}\!=\!0$) in these relationships yields the result in Fig.~\ref{Fig:ConceptofDBM}(e) for a dBM wave packet associated with a particle in its rest frame.

\begin{figure*}[t!]
\centering
\includegraphics[width=17.6cm]{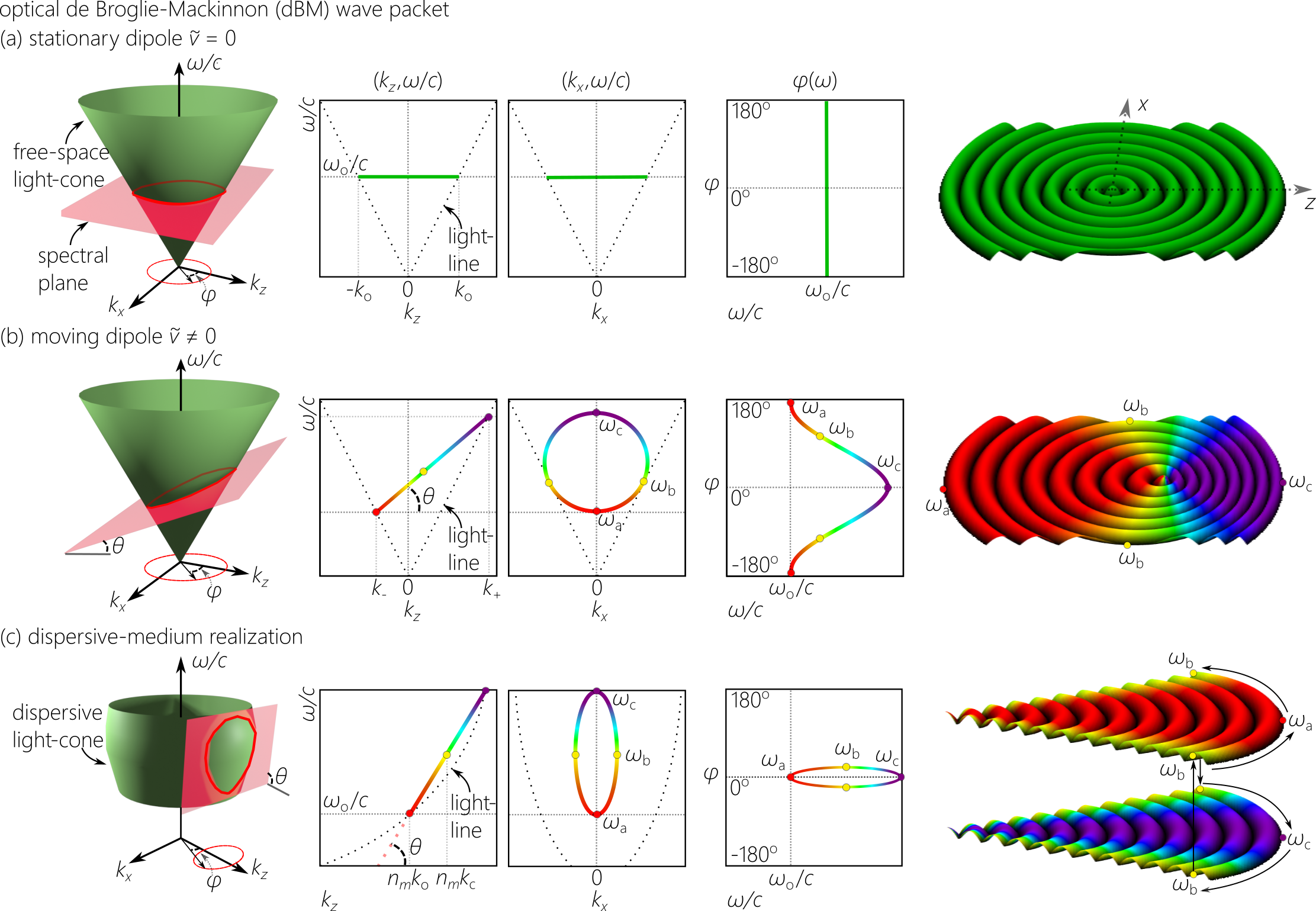}
\caption{\textbf{Optical de~Broglie-Mackinnon (dBM) wave packets.} From left to right we depict the following: the light-cone in $(k_{x},k_{z},\tfrac{\omega}{c})$-space intersecting with a spectral constraint in the form of a plane; the spectral projection onto the $(k_{z},\tfrac{\omega}{c})$-plane; the spectral projection onto the $(k_{x},\tfrac{\omega}{c})$-plane; the propagation angle $\varphi(\omega)$; and the real part of the spatio-temporal field profile $\psi(x,z;t)$ at a fixed axial plane $z$. (a) A stationary monochromatic planar dipole in free space resulting from the constraint $\omega\!=\!\omega_{\mathrm{o}}$. (b) A moving planar dipole in free space corresponding to the constraint $k_{z}\!=\!\tfrac{1}{\beta_{v}}(\tfrac{\omega}{c}-\tfrac{k_{\mathrm{o}}^{2}}{k_{1}})$. (c) The field in a dispersive medium in the anomalous regime after imposing the constraint $k_{z}\!=\!\tfrac{1}{\beta_{v}}\{\tfrac{\omega}{c}-k_{\mathrm{o}}(1-n_{\mathrm{m}}\beta_{v})\}$.}
\label{Fig:OpticaldBMConcept}
\end{figure*}

\noindent\textbf{Optical dBM wave packets in free space.} For a monochromatic optical field at $\omega\!=\!\omega_{\mathrm{o}}$, we have the dispersion relationship $k_{x}^{2}+k_{z}^{2}\!=\!k_{\mathrm{o}}^{2}$, which is the circle at the intersection of the free-space light-cone $k_{x}^{2}+k_{z}^{2}\!=\!(\tfrac{\omega}{c})^{2}$ with the horizontal iso-frequency plane $\omega\!=\!\omega_{\mathrm{o}}$. As described in the main text, the full circle is the spectral support for the optical field produced by a stationary planar dipole. When the source and detector are in relative motion along the $z$-axis at a velocity $v$, the initially horizontal iso-frequency-plane in $(k_{x},k_{z},\tfrac{\omega}{c})$-space is tilted by an angle $\theta$ with respect to the $k_{z}$-axis,where $\tan{\theta}\!=\!\beta_{v}$ \cite{Kondakci18PRL}. The resulting spectral constraint at the intersection with the light-cone (that conforms to a dBM wave packet) is:
\begin{equation}
k_{z}=k_{+}+\frac{\omega-\omega_{+}}{\widetilde{v}}=\frac{1}{\beta_{v}}\left(\frac{\omega}{c}-k_{\mathrm{o}}\sqrt{1-\beta_{v}^{2}}\right)=\frac{1}{\beta_{v}}\left(\frac{\omega}{c}-\frac{k_{\mathrm{o}}^{2}}{k_{1}}\right),
\end{equation}
where $k_{+}\!=\!\tfrac{\omega_{+}}{c}$ is the point on the light-line $k_{z}\!=\!\tfrac{\omega}{c}$ intersecting with the tilted spectral plane, and we make use of the same definitions of $k_{\mathrm{o}}$, $k_{+}$, and $k_{1}$ as above for dBM wave packets. Eliminating $k_{z}$ from these two relationships (the light-cone and the titled spectral plane) yields the spectral projection onto the $(k_{x},\tfrac{\omega}{c})$-plane in the form of an ellipse (Eq.~\ref{Eq:EllipseInFreeSpace}) [Fig.~\ref{Fig:OpticaldBMConcept}(b)]. We can also obtain the spectral projection onto the $(k_{x},k_{z})$-plane by eliminating $\omega$: substituting $\tfrac{\omega}{c}\!=\!\beta_{v}k_{z}+k_{\mathrm{o}}\sqrt{1-\beta_{v}^{2}}$ from the spectral constraint into $k_{x}^{2}+k_{z}^{2}\!=\!(\tfrac{\omega}{c})^{2}$ yields the ellipse $\tfrac{k_{x}^{2}}{k_{\mathrm{o}}^{2}}+\tfrac{1}{(\Delta k_{z}/2)^{2}}(k_{z}-\beta_{v}k_{1})^{2}\!=\!1$, where $\Delta k_{z}\!=\!2k_{1}$, and $\beta_{v}k_{1}$ is the central axial wave number. Substituting $k_{x}\!=\!\tfrac{\omega}{c}\sin{\varphi}$ and $k_{z}\!=\!\tfrac{\omega}{c}\cos{\varphi}$, the propagation angle is $\cos{\varphi(\omega)}\!=\!\{1-\tfrac{\omega_{\mathrm{o}}}{\omega}\sqrt{1-\beta^{2}}\}/\beta_{v}$, $\omega\!\neq\!0$, which is \textit{not} differentiable at $\varphi\!=\!0$ or $\varphi\!=\!\pi$ -- corresponding to the maximum and minimum points on the ellipses in the $(k_{x},\tfrac{\omega}{c})$ or $(k_{x},k_{z})$ planes.

\noindent\textbf{Optical dBM wave packets in presence of anomalous GVD.} In presence of anomalous GVD, the wave number in the medium expanded around $\omega_{\mathrm{o}}$ is $k\!=\!n(\omega)\omega/c\!=\!n_{\mathrm{m}}k_{\mathrm{o}}+\Omega/\widetilde{v}_{\mathrm{m}}-\tfrac{1}{2}|k_{2\mathrm{m}}|\Omega^{2}$; here $\Omega\!=\!\omega-\omega_{\mathrm{o}}$ and $n(\omega)$ is the frequency-dependent refractive index. The quantities $n_{\mathrm{m}}$, $\widetilde{v}_{\mathrm{m}}$, and $k_{2\mathrm{m}}$ are all evaluated in the medium at $\omega\!=\!\omega_{\mathrm{o}}$: $n_{\mathrm{m}}\!=\!n(\omega_{\mathrm{o}})$ is the refractive index; $\widetilde{v}\!=\!1\big/\tfrac{dk}{d\omega}\big|_{\omega_{\mathrm{o}}}$ is the group velocity, and $k_{2\mathrm{m}}\!=\!\tfrac{d^{2}k}{d\omega^{2}}\big|_{\omega_{\mathrm{o}}}\!=\!-|k_{2\mathrm{m}}|$ is the negative-valued GVD coefficient in the anomalous dispersion regime. In the small-angle (paraxial) approximation,
\begin{equation}\label{Eq:kzInDispersiveMedium}
k_{z}=\sqrt{k^{2}-k_{x}^{2}}\approx n_{\mathrm{m}}k_{\mathrm{o}}+\frac{\Omega}{\widetilde{v}_{\mathrm{m}}}-\frac{1}{2}|k_{2\mathrm{m}}|\Omega^{2}-\frac{k_{x}^{2}}{2n_{\mathrm{m}}k_{\mathrm{o}}},
\end{equation}
and the light-line $k_{x}\!=\!0$) is now curved [Fig.~\ref{Fig:OpticaldBMConcept}(c)]. To produce a dBM wave packet, we impose the spectral constraint
\begin{equation}\label{Eq:kzConstraintInDispersiveMedium}
k_{z}=n_{\mathrm{m}}k_{\mathrm{o}}+\frac{\Omega}{\widetilde{v}},
\end{equation}
which intersects with the light-line at two points: $\Omega\!=\!0$ ($\omega\!=\!\omega_{\mathrm{o}}$) and $\tfrac{\omega}{c}\!=\!-2k_{\mathrm{a}}\tfrac{1-\beta_{v}'}{\beta_{v}'}$; here $\beta_{v}'\!=\!\tfrac{\beta_{v}}{\beta_{\mathrm{m}}}\!=\!\tfrac{\widetilde{v}}{\widetilde{v}_{\mathrm{m}}}$, $\widetilde{v}\!<\!\widetilde{v}_{\mathrm{m}}$, $k_{\mathrm{a}}\!=\!(c^{2}|k_{2\mathrm{m}}|\beta_{\mathrm{m}})^{-1}\!=\!\tfrac{\omega_{\mathrm{a}}}{c}$. The maximum temporal bandwidth compatible with dispersion-free propagation in this dispersive medium at a group velocity $\widetilde{v}$ is thus $\tfrac{\Delta\omega}{c}\!=\!2k_{\mathrm{a}}\tfrac{1-\beta_{v}'}{\beta_{v}'}$, and the corresponding maximum bandwidth of the axial wave number is $\Delta k_{z}\!=\!2\tfrac{k_{\mathrm{a}}}{\beta_{\mathrm{m}}}\tfrac{1-\beta_{v}'}{\beta_{v}'^{2}}$, so that $\tfrac{\Delta\omega/c}{\Delta k_{z}}\!=\!\beta_{\mathrm{m}}\beta_{v}'\!=\!\beta_{v}$.

We can now obtain the spectral projections onto the $(k_{x},\tfrac{\omega}{c})$ and $(k_{z},\tfrac{\omega}{c})$ planes for the optical dBM wave packet in presence of anomalous GVD just as we did for their counterparts in free space. By equating Eq.~\ref{Eq:kzInDispersiveMedium} and Eq.~\ref{Eq:kzConstraintInDispersiveMedium} we eliminate $k_{z}$ and obtain in the $(k_{x},\tfrac{\omega}{c})$-plane the ellipse $\tfrac{k_{x}^{2}}{k_{x,\mathrm{max}}^{2}}+\tfrac{(\omega-\omega_{\mathrm{c}})^{2}}{(\Delta\omega/2)^{2}}\!=\!1$, where the central frequency is $\omega_{\mathrm{c}}\!=\!\omega_{\mathrm{o}}-\tfrac{\Delta\omega}{2}$, $k_{x,\mathrm{max}}^{2}\!=\!\sigma_{\mathrm{m}}(\tfrac{\Delta\omega/2}{c})^{2}$, and $\sigma_{\mathrm{m}}\!=\!n_{\mathrm{m}}c\omega_{\mathrm{o}}|k_{2\mathrm{m}}|$ is a dimensionless GVD parameter. Similarly, we can obtain the spectral projection onto the $(k_{x},k_{z})$-plane by substituting $\Omega\!=\!\widetilde{v}(k_{z}-n_{\mathrm{m}}k_{\mathrm{o}})$ from the spectral constraint in Eq.~\ref{Eq:kzConstraintInDispersiveMedium} into the light-cone in Eq.~\ref{Eq:kzInDispersiveMedium} to obtain the ellipse: $\tfrac{k_{x}^{2}}{k_{x,\mathrm{max}}^{2}}+\tfrac{(k_{z}-k_{\mathrm{c}})^{2}}{(\Delta k_{z}/2)^{2}}\!=\!1$, where $k_{\mathrm{c}}$ is the center of the $k_{z}$-span.

\begin{figure*}[t!]
\centering
\includegraphics[width=17cm]{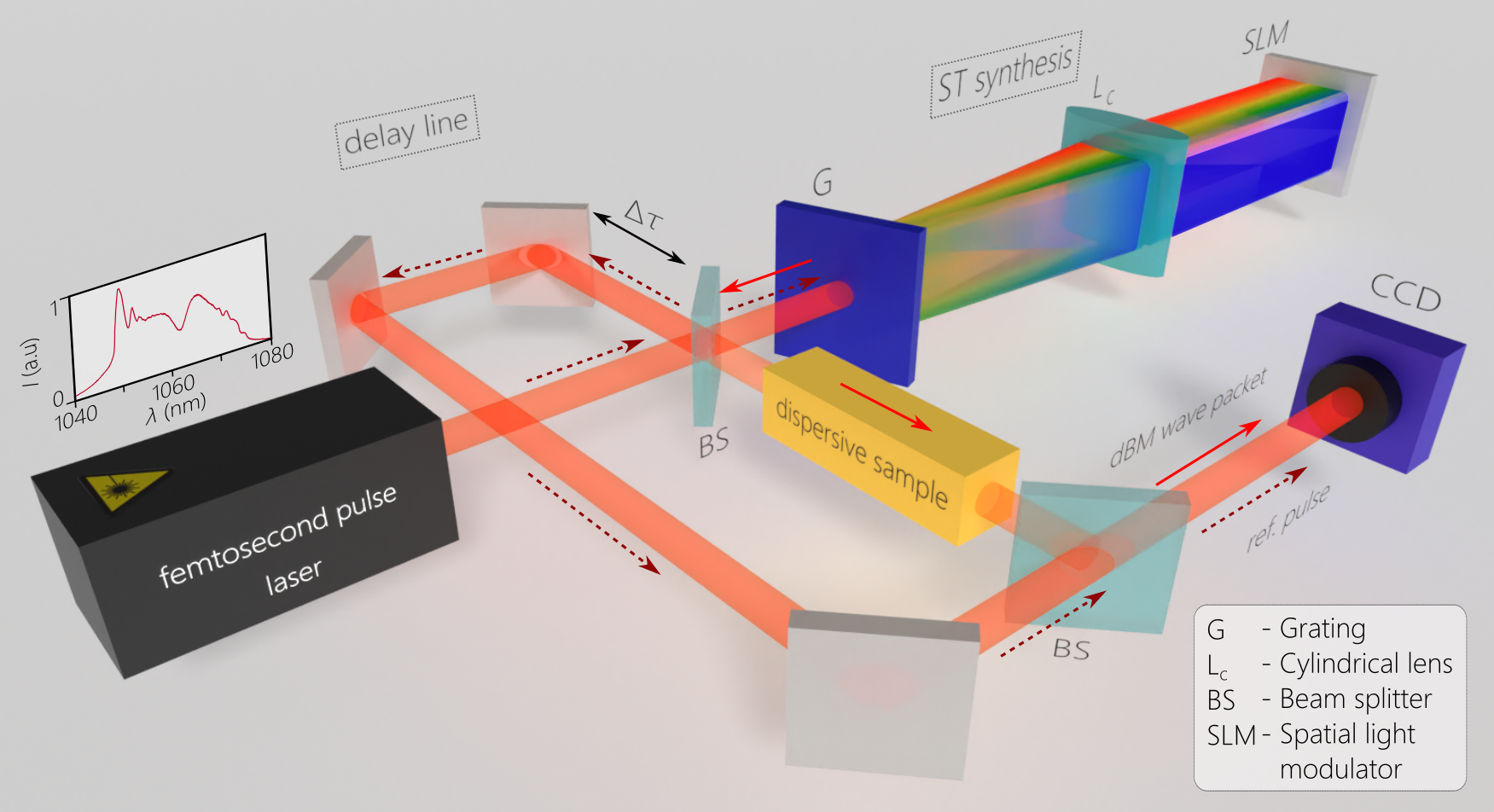}
\caption{\textbf{Synthesizing and characterizing optical dBM wave packets.} Starting with femtosecond laser pulses, the space-time (ST) synthesis arrangement associates each wavelength $\lambda$ with prescribed propagation angles $\pm\varphi(\lambda)$ before traversing the dispersive medium \cite{Hall21OE2}. Interfering the synthesized optical dBM wave packets with reference plane-wave pulses from the initial laser helps reconstruct the spatio-temporal intensity profile of the dBM wave packets.}
\label{Fig:Setup}
\end{figure*}

\noindent\textbf{Details of the experimental setup and spectral measurements.} The field configuration shown in Fig.~\ref{Fig:OpticaldBMConcept}(c) is produced via the angular-dispersion synthesizer depicted in Fig.~\ref{Fig:Setup} []. Starting with femtosecond laser pulses (central wavelength $\lambda\!=\!1064$~nm, bandwidth $\Delta\lambda\!=\!20$~nm, and pulsewidth $\Delta T\!\approx\!100$~fs; Spark Lasers, Alcor), a diffraction grating (1200~lines/mm) resolves the spectrum spatially, and a cylindrical lens (focal length $f\!=\!500$~mm) collimates the wave front before incidence on a reflective, phase-only SLM (Meadowlark, E19X12). The SLM deflects each wavelength $\lambda$ at angles $\pm\varphi(\lambda)$ according to the elliptical spatio-temporal spectrum given in Eq.~\ref{Eq:DispersiveEllipseOmegaKx}. The wave front retro-reflected from the SLM returns to the grating, and the optical dBM wave packet is formed. The spatio-temporal spectrum is acquired by directing a portion of the spectrally resolved wave front reflecting back from the SLM to a spatial Fourier transform (not shown in Fig.~\ref{Fig:Setup} for simplicity), which thus yields the spatio-temporal spectral projection onto the $(k_{x},\lambda)$-plane. We then obtain the spectral projection onto the $(k_{z},\lambda)$-plane in the medium via the relationship $k_{z}(\omega)\!=\!\sqrt{\{(n(\omega)\tfrac{\omega}{c}\}^{2}-k_{x}^{2}(\omega)}$.

\noindent\textbf{Dispersive medium.} The dispersive sample that we exploit comprises a pair of chirped Bragg mirrors (Edmund 12-335) that provide anomalous group-delay dispersion (GDD). By adjusting the separation  between the two mirrors and the incident angle of the wave packet, we can control the number of bounces off the mirrors, thereby producing an anomalous-GVD coefficient of $k_{2\mathrm{m}}\!\approx\!-500$~fs$^{2}$/mm. Because the thickness of the mirrors is negligible with respect to the free-space gap separating them, we can thus have $\widetilde{v}_{\mathrm{m}}\!\approx\!c$ ($\widetilde{n}_{\mathrm{m}}\!\approx\!1$).

\noindent\textbf{Reconstruction of the spatio-temporal profiles of dBM wave packets.} To reconstruct the spatio-temporal intensity profile of a dBM wave packet $I(x,z;\tau)$ at a fixed axial plane $z$, we make use of the interferometric arrangement depicted schematically in Fig.~\ref{Fig:Setup}. We bring together two wave packets: the synthesized optical dBM wave packet, and a reference plane-wave pulse taken from the initial laser pulse \cite{Kondakci19NC}. An optical delay $\tau$ is placed in the path of the reference pulse. When the dBM wave packet and the reference pulse overlap in space and time, we observe high-visibility spatially resolved interference fringes at the CCD camera placed in their common path. As we sweep the optical delay $\tau$ (thus reducing the overlap between the dBM wave packet and the reference pulse), the interference visibility drops. We make use of the recorded visibility along $x$ and $tau$ to reconstruct the wave packet intensity profile $I(x;\tau)$.

\clearpage

\end{document}